# Superpower Glass: Delivering Unobtrusive Real-time Social Cues in Wearable Systems

**Catalin Voss** — catalin@cs.stanford.edu
**Peter Washington** — peterwashington@stanford.edu
**Nick Haber** — nhaber@stanford.edu
**Aaron Kline** — akline@stanford.edu
**Jena Daniels** — danielsj@stanford.edu
**Azar Fazel** — azarf@stanford.edu
**Titas De** — titasde@stanford.edu
**Beth McCarthy** — bethmac@stanford.edu
**Carl Feinstein** — carlf@stanford.edu
**Terry Winograd** — winograd@cs.stanford.edu
**Dennis Wall** — dpwall@stanford.edu

Stanford University
Stanford, CA 94305, USA

## Abstract
We have developed a system for automatic facial expression recognition, which runs on Google Glass and delivers real-time social cues to the wearer. We evaluate the system as a behavioral aid for children with Autism Spectrum Disorder (ASD), who can greatly benefit from real-time non-invasive emotional cues and are more sensitive to sensory input than neurotypically developing children. In addition, we present a mobile application that enables users of the wearable aid to review their videos along with auto-curated emotional information on the video playback bar. This integrates our learning aid into the context of behavioral therapy. Expanding on our previous work describing in-lab trials, this paper presents our system and application-level design decisions in depth as well as the interface learnings gathered during the use of the system by multiple children with ASD in an at-home iterative trial.

## ACM Classification Keywords
H.5.m. Information interfaces and presentation (e.g., HCI): Miscellaneous.

**Introduction**

This paper describes a system we have built, named SuperpowerGlass, for automatic facial expression recognition, which runs on Google Glass and delivers real-time social cues to users. The system tracks expressive events in faces using the outward-facing camera on the device. It then provides social cues to the wearer, shown on a heads-up display or as audio cues. SuperpowerGlass can also record social responses from the user, such as the amount and type of eye contact using Glass' sensors and a custom-made infrared eye tracker.

An important application of our system is as a behavioral aid for children with developmental disorders, as these individuals almost unanimously struggle with social interactions. Developing a system to teach cognitively impaired children social engagement skills presents several design challenges. Because children with many mental disorders including Autism Spectrum Disorder (ASD) can hyper-focus on nonsocial objects and react negatively to sensory overload [26], the system must not distract children from the primary objective of increasing social awareness and abilities.

SuperpowerGlass has the potential to significantly enhance behavioral therapy. The current standard of care for ASD and some other cognitive disorders consists of "flashcard therapy" involving painstaking memorization of facial emotions [21]. While computer-assisted in-place treatment systems have been studied for years [3, 15, 39], few strides have been made to bring the learning process of social interactivity away from the clinician's office or flashcards and into daily life of patients. Previous attempts to design just-in-time in-situ learning aids for facial affect showed promising potential by presenting emotional cues in a fun and intuitive way on a mini-computer [25]. Such systems did not achieve widespread use due to the obtrusive nature of the feedback aids and the state of expression recognition technology available at the time.

Our system supports multiple tasks to be integrated into therapy for cognitively impaired children. To provide engaging interaction methods, we include various gamified activities that the children can use during informal behavioral therapy sessions. When the glasses are worn at home, they can be used as an unstructured emotional aid during social activities (e.g., dinner). To integrate our primary system for delivering unobtrusive real-time social cues into the context of behavioral therapy, we have developed an Android application that allows parents and children to review activities recorded throughout the day.

"Emotional moments" are auto-curated and highlighted in the video playback mode, as shown in Figure 5. This allows users to skip to emotionally relevant parts of the video, crucial for longer videos that are difficult to navigate manually. This review system can serve as an effective additional therapeutic tool for children with ASD, as it provides the opportunity for meaningful discussion of the emotions and mental states of the child and their family members, as captured in the context in which they occurred. Parents are encouraged to review these moments with their children (and if they choose, behavioral therapists).

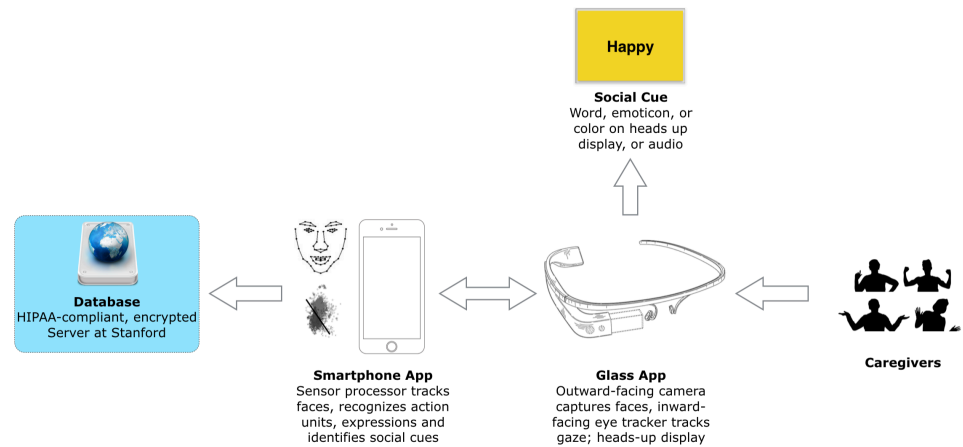

Figure 1: An overview of the SuperpowerGlass system.

This paper presents the system architecture of SuperpowerGlass and then discusses how results from an in-lab study with 20 children and early observations from an at-home trial with 12 children (scaling up to 100 participants) with Autism between 4 and 17 years old prompted key design choices in developing our wearable aid. We conclude by discussing tradeoffs between the design choices we considered and possible lessons for other wearable systems that provide real-time cues during social interaction.

## Related Work

Here, we review work on ways to provide effective social cues in wearable systems, quick video browsing, and interaction techniques for people with disabilities.

*Social Cues on Wearable Vision Systems*
There have been several studies exploring the interface for wearable systems that perform vision tasks and provide cues to the user. Techniques for providing AR social cues to enhance a user's understanding of the environment have been studied in a variety of contexts. VizWear runs vision tasks in real-time and provides a graphical overlay that displays 2-dimensional annotations based on the wearer's view [20]. GravitySpot guides users in front of public displays using on-screen social cues [1]. The wearable eye-q delivers discrete visual cues to the user; the user study found that the cues can be designed to meet specific levels of disruption for the wearer, so that some cues are less noticeable when the user is not under high workload [4]. SWAN is a system that provides wearable *audio* navigation to the visually impaired [41].

Developing energy-efficient systems that perform continuous mobile vision tasks is difficult because modern image sensors are not energy-proportional [22]. To deal with the low battery life of wearable devices, heavy computation is usually offloaded to the cloud or a remote server [19]. We follow a similar computational offloading model in our system, using a locally-linked Android phone to achieve the low local-network latency required in our application.

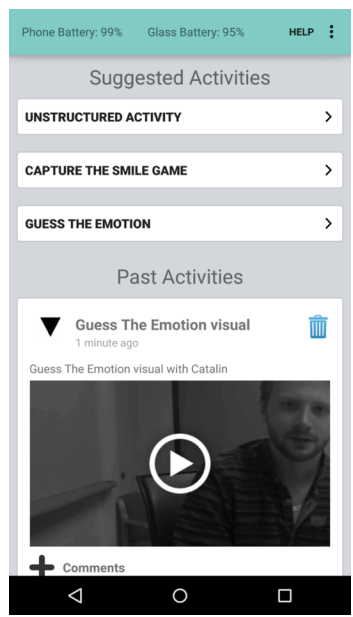

Figure 2: Home screen of the android phone application, showcasing the caregiver review. From here, study participants can start activities and review data from past sessions.

*Quick Video Browsing*
There has been work on effectively providing video summaries to users in the form of video digests that afford browsing and skimming the videos through segmenting the video into cohesive subdivisions [29]. GliFlix is a system that uses augmented subtitles to selectively highlight foreign words spoken in the films, together with the users' primary-language subtitles [33]. Retrieving emotional content from videos has received little attention, and techniques focus on having users label emotions explicitly [2].

*Interaction Techniques for Cognitive Disabilities*
Interaction techniques to aid in therapy for cognitive disabilities have been studied as well. VibroGlove is an assistive aid for conveying facial expressions via haptic interactions [18]. With the advent of mobile and wearable computing, researchers have begun to explore the use of pervasive computing to aid in the care of children with ASD. Picard's lab at MIT has explored the use of wearable systems as a companion tool for social-emotional learning and the use of the recorded videos for defining a process to collect, segment, label, and use video clips from everyday conversations [38]. Kientz's work has determined several relevant principles for designing ubiquitous systems for children with ASD: changes in behavior should be seamless for the child, easier systems are better, customization of the system is critical, and parents are usually a better source of feedback about the system than the children themselves [17]. These design principles relate well to our system and study.

Watching a television series designed to enhance emotion comprehension every day can improve the comprehension of young children with ASD [10]. Multitouch interfaces for behavioral therapy and have been found to increase collaboration skills in children with ASD [2, 8, 9]. VR immersive headsets have also been found to be effective learning aids for both children and young adults [16, 36]. Humanoid robots used as catalysts for social behavior in ASD therapy have been shown to promote positive interactions [7, 11]. Much of this work has determined that children with ASD enjoy gamified systems and that a playful system is generally most effective in providing useful behavioral therapy to children with ASD. The work of [28] designs games to teach emotion to children with ASD. Techniques from affective computing have recently been found useful in providing an effective aid in behavioral therapy for children with ASD [30, 31, 32]. Picard's group has identified challenges of using Google Glass to detect emotions, such as a constrained battery life, limited storage space, and privacy concerns over transmitting affective information wirelessly [13].

## Architecture
Here, we describe the various system architecture decisions made in developing SuperpowerGlass.

*Communication Between Glass and Android*
The glasses and Android device connect via a dedicated wireless network, originating from the Android device and secured with WPA2 encryption. Users initiate a new behavioral-therapy activity through the Android application, prompting Glass to begin the activity and start capturing video frames. To minimize latency, Glass then sends the video data to the phone via UDP as uncompressed greyscale frames at 320x194 resolution at a frame rate of 30 fps, along with control data and metadata. The phone provides a live preview of the incoming video feed, encodes and stores the video data, and runs the frames through an expression recognition pipeline. It then returns detected emotions, face location data, and user command requests via UDP to the glasses. Glass interprets the result and shows the appropriate social cue (either on the heads-up display or as audio); it also handles command requests and uses the facial location data to inform the subsequent video data transmission (described below).

*Increasing Facial Resolution*
The combination of camera processing and network activity on the glasses still results in a limitation on the resolution of images sent at a frame rate sufficient for the emotion recognition processor. This limitation increases over time as the device temperature rises, resulting in CPU throttling. In order to balance frame rate and resolution needs, a dual approach is taken. A lower resolution full-frame image is provided whenever a face has not been detected; this allows for face detection through the camera's full field of view. When a face has been detected, the face's location in the field of view is used to provide a cropped view of the face from a higher resolution image. A simple algorithm on the glasses iteratively computes an optimal face crop from a series of current face estimates coming from the phone to meet a number of simultaneous objectives: a) hold the frame steady to avoid sudden jumps that decrease the viewability of the resulting video or disturb the emotion recognition pipeline, b) keep the face centered in the frame, and c) correct for small errors in the face detector or face tracker estimates of face bounds. A series of moving average filters smooth these estimates.

*Emotion Recognition Models*
Our expression recognition system employs SVM and logistic regression classifiers trained on HOG features [5] after face tracker registration and lighting normalization [37]. We use a mix of academic databases [4, 12, 23, 24, 27, 34, 35] along with our own data gathered in-lab, to train these classifiers. The system employs a method termed *neutral subtraction* that learns the subject's neutral face at runtime in order to better discriminate expressions [13]. The system is modular in that the facial expression recognition classifiers are trained separately from other components of the computer vision system. This allows us not only to quickly evaluate the experience with different models – evaluated via A/B testing – but also to adapt the system to different face-relevant tasks. We filter the results of these classifiers to reduce the fire-rate of social cues and avoid false positives or rapid switching between related cues.

*Emotion Recognition Training Pipeline*
For the purpose of behavioral intervention, it is imperative that expression recognition performs optimally on specific individuals – that is, generalized high accuracy is secondary to ideal behavior when viewing members of the child's family. We have built systems by which recognition can be calibrated rapidly on selected individuals. When introducing the family to the device, we use an iterative training technique in which we ask subjects to act out various expressions after which we adapt the expression recognition model and let an iterative learning algorithm decide what further acted expressions are needed, if any. This gives real-time feedback to determine when the expression recognition system converges to optimal performance in the child's natural environment.

---

**Visual Feedback Cues**

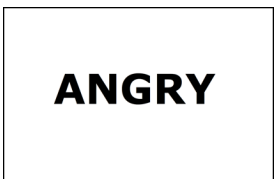

**Text**: The name of the emotion is written on the screen. The screen of the Google Glass contains a white background and the font color is black.

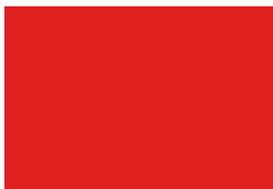

**Color**: Work in color psychology suggests that humans associate colors with certain emotions. In color mode, the screen is filled with the color associated with the recognized emotion.

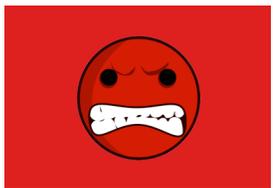

**Smiley**: A playful smiley representing the emotion is displayed on the screen.

Figure 3: Visual feedback cues.

## Real-Time Feedback Mechanisms

An ideal feedback mechanism will maximize behavioral information that the child receives while minimizing the distractions to the user. The feedback can primarily come in the form of audio or visual cues. As shown in Figure 3, we explored a variety of visual feedback cues.

*Visual Feedback*

We tested three primary emotional visual feedback mechanisms: text, color, and emoticons. The visual feedback can be any combination of these 3 basic feedback mechanisms. However, the number of visual cues provided is kept to a minimum because children with ASD can experience sensory overload [26].

*Audio Feedback*

The audio feedback includes a narrator reading out the name of the emotion as well as a range of child-friendly playful sound effects associated with the emotion. In addition to the possibilities for visual and audio feedback individually, the cues can also be combined, allowing for a large range of feedback mechanisms.

*Face-Tracking Indicator*

During the in-lab trials conducted in [40], we observed situations in which the system would fail to recognize emotions not because of a problem with the recognition model but because a face was not detected, for example due to challenging backlighting. To enable users to distinguish between a lack of cues and a correctly tracked yet neutral face, we added an indicator signal on the display of the glasses that lights up whenever a face is detected, as shown in Figure 4. This may introduce a problem, potentially encouraging the child to view the indicator overlaid on their conversation partner. In order to attempt to combat these possibilities, we have designed several possible indicators, as detailed in Figure 4, tested and discussed in the User Study section.

## Caregiver Review

The Android phone serves as a caregiver review system for previous sessions. The review system contains a newsfeed-like view of the previous session recordings in chronological order, as shown in Figure 2. For each video, "emotional moments" are auto-curated and highlighted in the video playback mode, as shown in Figure 5. This allows users to skip to emotionally-relevant parts of the video, which is crucial for longer videos. Caregivers can further tag particular parts of video as important, hide or permanently delete videos from the newsfeed, and add comments to the video. These options serve as tools for the parents and feedback mechanisms for researchers.

All parent interaction with the caregiver review system save to the SD card of the Android device so that researchers can study interactions with the system.

## At-Home User Study

Following a more constrained in-lab study with 40 participants (20 children with and 20 without ASD) described in [40], we conducted a 4-month at-home iterative design trial exclusively with ASD children in order to test the various parameters of our system such as the emotion recognition model, feedback mechanisms, and indicator design. We recruited 12 families, who we asked to take the system home and to use for one hour every day. Every 2 weeks, the families visited the lab, where we uploaded their videos, comments, and marked sections from the SD card on the Android to our secure HIPAA-compliant server. We

**Indicators**

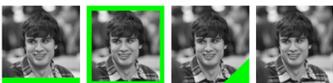

Figure 4: Face tracking indicator options tested with children during the at-home iterative design trials.

**Caregiver Review**

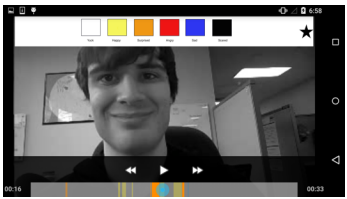

Figure 5: While reviewing the videos, parents have the ability to mark particular parts of the session as important. This can signify a variety of things to both the parents and researchers, such as an egregious error in emotion recognition, interesting behavior by the child in response to the behavioral therapy, or an otherwise important part of the video that should be reviewed at a later time.

then provided the parents with an updated version of the glasses and Android apps. This allowed us to test the effectiveness of various parameters of our system, such as the design of the face detection indicator and the emotion recognition model used. Based on a review of the videos during the 2-week check-ins, we made several observations about the use of the system outside of the lab, giving us significant insights about the effectiveness and usability of the system.

Behavioral Findings
*Parents noticed an improvement in child eye contact from using the system.* After a few days of use, parents commented that their children made more eye contact during conversation, even when not wearing SuperpowerGlass. One child's teacher, blind to the study, remarked on the increase in the child's eye contact in class to the child's parent.

*The participants greatly enjoy conducting the activities.* The children reported positive experiences with the activities at home. They stated that they viewed the system as a toy. However, interest in the system steadily declined over the period that the children participated in the trial.

*The participants stop use if the device if becomes too warm.* Children will comment that the device is uncomfortable after it has been running for a long time. To combat this, we told parents to limit the time of any given activity session to 30 minutes at a time, giving the glasses time to cool down before the next session.

Feedback Mechanism Findings
The optimal feedback mechanism is a combination of visual and audio feedback. Contrary to our concerns about sensory overload with children with ASD and to our findings from our in-lab study [40], the combination of audio and visual feedback was not too distracting to the children.

Indicator Findings
We observed some participants tilting their heads (an unintended and potentially disruptive effect) in order to set off the indicator during a separate kickoff session with five neurotypical children. However, we saw no evidence that participants with ASD replicated this behavior at home based on our video material and parent interviews. Children displayed a strong preference for the box indicator, contrary to our hypothesis that they would prefer a less visually obtrusive indicator. They were more accepting of the line indicator and disliked the triangle indicator, stating that they found it distracting.

## Conclusions and Future Work
Our initial findings have shown that SuperpowerGlass can be an effective tool for delivering real-time emotion cues to a user, particularly for children with ASD. We gained key insights about building such a system, namely that children with ASD respond optimally to a combination of visual and audio feedback despite concern about the negative effects of sensory overload.

We would like to explore ways that the emotion recognition models can be further improved through iterative training at home. We plan to build a mobile interface that allows parents to quickly train the emotion recognition model on themselves in instances when certain emotions are frequently misclassified. This would provide an interface that allows subjects to improve the model themselves (and ensure that they do not cause undesirable behavior). Although the model should theoretically converge after a few iterations, designing an interface that makes providing repetitive training data to a machine learning model easy, intuitive, and even fun, remains an open HCI research area.

We would also like to explore and evaluate alternative games and activities that can be used with our system, including games that can subtly update the active learning model. These activities would enable participants to label emotions through the phone as they saw them. The system can use these labels as additional data, adapting to the natural conditions of the subjects' homes.

Finally, we plan to expand our study from an iterative design trial into a formal systematic study that measures quantitative improvements in emotional intelligence in 100 children with ASD.

## Acknowledgments

This work was supported with funding from the Dekeyser and Friends Foundation and the David and Lucile Packard Foundation under Special Projects Grant 2015-62349. We would also like to thank Google for donating 35 units of Google Glass.